\documentstyle[12pt,epsf,epsfig]{article}
\newcount\timecount
\newcount\hours \newcount\minutes  \newcount\temp \newcount\pmhours

\hours = \time
\divide\hours by 60
\temp = \hours
\multiply\temp by 60
\minutes = \time
\advance\minutes by -\temp
\def\hour{\the\hours}
\def\minute{\ifnum\minutes<10 0\the\minutes
            \else\the\minutes\fi}
\def\clock{
\ifnum\hours=0 12:\minute\ AM
\else\ifnum\hours<12 \hour:\minute\ AM
      \else\ifnum\hours=12 12:\minute\ PM
            \else\ifnum\hours>12
                 \pmhours=\hours
                 \advance\pmhours by -12
                 \the\pmhours:\minute\ PM
                 \fi
            \fi
      \fi
\fi
}

\def\monthname{\relax\ifcase\month 0/\or January\or February\or
   March\or April\or May\or June\or July\or August\or September\or
   October\or November\or December\else\number\month/\fi}

\def\bold#1{\setbox0=\hbox{$#1$}%
     \kern-.025em\copy0\kern-\wd0
     \kern.05em\copy0\kern-\wd0
     \kern-.025em\raise.0433em\box0 }

\def\ga{\mathrel{\raise.3ex\hbox{$>$\kern-.75em\lower1ex\hbox{$\sim$}}}}
\def\la{\mathrel{\raise.3ex\hbox{$<$\kern-.75em\lower1ex\hbox{$\sim$}}}}
\def\gev{{\rm \, Ge\kern-0.125em V}}
\def\tev{{\rm \, Te\kern-0.125em V}}
\def\beq{\begin{equation}}
\def\eeq{\end{equation}}

\def\ohsq{\Omega_{\chi} h^2}

\def\m12{m_{1\!/2}}

\begin{document}
\begin{titlepage}
\pagestyle{empty}
\baselineskip=21pt
\rightline{hep-ph/0111064}
\rightline{CERN--TH/2001-291}
\rightline{UMN--TH--2029/01}
\rightline{TPI--MINN--01/47}
\vskip 0.2in
\begin{center}
{\large{\bf Constraints from Accelerator Experiments on the Elastic 
Scattering of CMSSM Dark Matter}}
\end{center}
\begin{center}
\vskip 0.2in
{{\bf John Ellis}$^1$, {\bf Andrew Ferstl}$^2$ and {\bf Keith
A.~Olive}$^{3}$}\\
\vskip 0.1in
{\it
$^1${TH Division, CERN, Geneva, Switzerland}\\
$^2${Department of Physics,
Winona State University, Winona, MN 55987, USA}\\
$^3${Theoretical Physics Institute,
University of Minnesota, Minneapolis, MN 55455, USA}}\\
\vskip 0.2in
{\bf Abstract}
\end{center}
\baselineskip=18pt \noindent

We explore the allowed ranges of cross sections for the elastic scattering
of neutralinos $\chi$ on nucleons in the constrained minimal
supersymmetric extension of the Standard Model (CMSSM), in which scalar
and gaugino masses are each assumed to be universal at some input grand
unification scale. We extend previous calculations to larger $\tan \beta$
and investigate the limits imposed by the recent LEP lower limit on the
mass of the Higgs boson and by $b \to s \gamma$, and those suggested by
$g_\mu - 2$.  The Higgs limit and $b \to s \gamma$ provide upper limits on
the cross section, particularly at small and large $\tan \beta$,
respectively, and the value of $g_\mu - 2$ suggests a lower limit on the
cross section for $\mu > 0$.  The spin-independent nucleon cross section
is restricted to the range $6 \times 10^{-8}$~pb$ > \sigma_{SI} > 2 \times
10^{-10}$~pb for $\mu > 0$, and the spin-dependent nucleon cross 
section to the range $ 10^{-5}$~pb$ > \sigma_{SD} > 2 \times 
10^{-7}$~pb. Lower values are allowed if $\mu <0$.

\vfill
\leftline{CERN--TH/2001-291}
\leftline{October 2001}
\end{titlepage}
\baselineskip=18pt

One of the front-running candidates for cold dark matter is the lightest
supersymmetric particle (LSP), which is often taken to be the lightest
neutralino $\chi$~\cite{EHNOS}. Several experiments looking for the
scattering of cold dark matter particles on nuclear targets~\cite{GW} have
reached a sensitivity to a spin-independent elastic cross section
$\sigma_{SI}$ of the order of $10^{-5}$~pb for $m_\chi \sim
100$~GeV~\cite{searches}, and one experiment has reported a possible
positive signal~\cite{DAMA}. A new generation of more sensitive
experiments is now being prepared and proposed, with sensitivities
extending as low as $3 \times 10^{-9}$~pb~\cite{GENIUS}. It is therefore
important to update theoretical predictions for the elastic
scattering cross section, including the spin-dependent component,
$\sigma_{SD}$, as well as the spin-independent part, $\sigma_{SI}$.

The cross-section ranges allowed in the general minimal supersymmetric
extension of the Standard Model (MSSM) are quite broad, being sensitive to
the Higgs and squark masses, in particular~\cite{etal,EFlO2}. It is common
to focus attention on the constrained MSSM (CMSSM), in which all the soft
supersymmetry-breaking scalar masses $m_0$ are required to be equal at an
input superysmmetric GUT scale, as are the gaugino masses $m_{1/2}$ and
the trilinear soft supersymmetry-breaking parameters $A$. These
assumptions yield well-defined relations between the various sparticle
masses, and correspondingly more definite predictions for the elastic
$\chi$-nucleon scattering cross sections as functions of
$m_\chi$~\cite{EFlO1}. This paper is devoted to an updated discussion of
$\sigma_{SI}$ and $\sigma_{SD}$ in the CMSSM as functions of $m_0$,
$m_{1/2}$, and $\tan \beta$ for $A = 0$.

This is timely in view of two significant experimental developments since
our previous analysis~\cite{EFlO2}. One has been the improvement in the 
experimental
lower limit from LEP on the mass of the lightest MSSM Higgs boson
$h$~\cite{LEPHiggs}, which is now $m_h > 114.1$~GeV in the context of the
CMSSM~\footnote{In the general MSSM, $m_h$ could be as low as $\sim
90$~GeV, but this is only possible for variants in which the $Z - Z - h$
coupling is suppressed to an extent that does not occur within the CMSSM
as studied here.}. The second major experimental development has been the
report of a possible 2.6 - $\sigma$ discrepancy between the measured and
Standard Model values of the anomalous magnetic moment of the muon, $a_\mu
\equiv (g_\mu - 2) / 2$:  $a_\mu = (43 \pm 16) \times
10^{-10}$~\cite{g-2}, which we interpret as $11 \times 10^{-10} < a_\mu <
75 \times 10^{-10}$. The supersymmetric interpretation~\cite{susyg-2,ENO}
of this result is not yet established: it could be that strong-interaction
uncertainties in the Standard Model prediction have been underestimated,
or there might have been a statistical fluctuation in the data. Even if
the discrepancy is confirmed, it might be evidence for some other type of
physics beyond the Standard Model. Nevertheless, we are tempted to explore
its possible consequences for dark matter scattering within the CMSSM
context~\cite{otherCMSSMDM}.

Theoretically, there have also been improvements recently in the
calculations in the CMSSM of the supersymmetric relic density $\Omega_\chi
h^2$ for large values of the ratio $\tan \beta$ of Higgs vacuum
expectation values~\cite{EFGOSi}. These define better the interesting
region of CMSSM parameter space where the relic density may fall within
the range $0.1 < \Omega_\chi h^2 < 0.3$ preferred by astrophysics and
cosmology~\cite{omegah2}.

We find that the expected ranges of both the spin-independent cross
sections $\sigma_{SI}$ and the spin-dependent cross sections $\sigma_{SD}$
in the CMSSM are quite restricted (see also~\cite{minmax}).  The LEP
Higgs limit~\cite{LEPHiggs} sets upper bounds on $\sigma_{SI}$ and 
$\sigma_{SD}$, not
only via the direct contribution of Higgs exchange to the scattering
matrix element, but also because it provides a strong lower limit on
$m_\chi$ at low $\tan \beta$, in particular~\cite{EGNO}.\footnote{Apart
from cancellations that occur in $\sigma_{SD}$ when $\mu < 0$, the
elastic cross sections are monotonically decreasing functions of $m_\chi$
in the CMSSM \cite{EFlO1}.} At high
$\tan
\beta$, the observed rate of $b
\to s \gamma$ also provides \cite{bsg} an important lower limit on
$m_\chi$ and hence an upper limit on $\sigma_{SI,SD}$ \cite{bsgt}. In
view of these upper limits, we are unable to provide a CMSSM
interpetation of the DAMA signal~\cite{DAMA}. More excitingly for
prospective experiments, the range $11 \times 10^{-10} < a_\mu < 75
\times 10^{-10}$ would imply important upper limits on sparticle masses,
and hence a {\it lower limit}: $\sigma_{SI} > 2 \times 10^{-10}$~pb.
Putting together all the constraints, we find for $\mu > 0$ a 
relatively
narrow band $6
\times 10^{-8}$~pb $ >
\sigma_{SI} > 2 \times 10^{-10}$~pb. The allowed range is typically
broadest at large $\tan \beta$.  Lower cross sections are  possible if
$\mu < 0$.

As has been discussed in detail elsewhere, the regions of $m_{1/2}, m_0$
plane where the relic density falls within the preferred range $0.1 <
\Omega_\chi h^2 < 0.3$ can be divided into four generic parts, whose
relative significances depend on $\tan \beta$. There is a `bulk' region at
moderate $m_{1/2}$ and $m_0$~\cite{EHNOS}, where supersymmetry is
relatively easy to detect at colliders and as dark matter. Then, extending
to larger $m_{1/2}$, there is a `tail' of the parameter space where the
LSP $\chi$ is almost degenerate with the next-to-lightest supersymmetric
particle (NLSP), and efficient coannihilations~\cite{oldcoann} keep
$\Omega_\chi h^2$ in the preferred range, even for larger values of
$m_\chi$~\cite{coann}. At larger $m_0$, close to the boundary where
electroweak symmetry breaking is no longer possible, there is the
`focus-point' region where the LSP has a more prominent Higgsino component
and
$m_\chi$ is small enough for $\Omega_\chi h^2$ to be
acceptable~\cite{focus}. Finally, extending to larger $m_{1/2}$ and $m_0$
at intermediate values of $m_{1/2} / m_0$, there may be a `funnel' of
CMSSM parameter space where rapid direct-channel annihilations via the
poles of the heavier Higgs bosons $A$ and $H$ may keep $\Omega_\chi h^2$
in the preferred range~\cite{EFGOSi,funnel}. In this paper, we focus on
the case
$A = 0$ and use the {\tt SSARD} code to calculate the relic
density~\cite{SSARD}.
The precise values of $m_{1/2}$ and $m_0$ in the `focus-point' and
`funnel' regions are quite sensitive to the precise values and treatments
of the input CMSSM and other parameters~\cite{EO,otherCMSSMomega}. These
regions are not emphasized in the following discussion, but are commented
upon where appropriate.

\begin{figure}
\vspace*{-0.75in}
\begin{minipage}{8in}
\epsfig{file=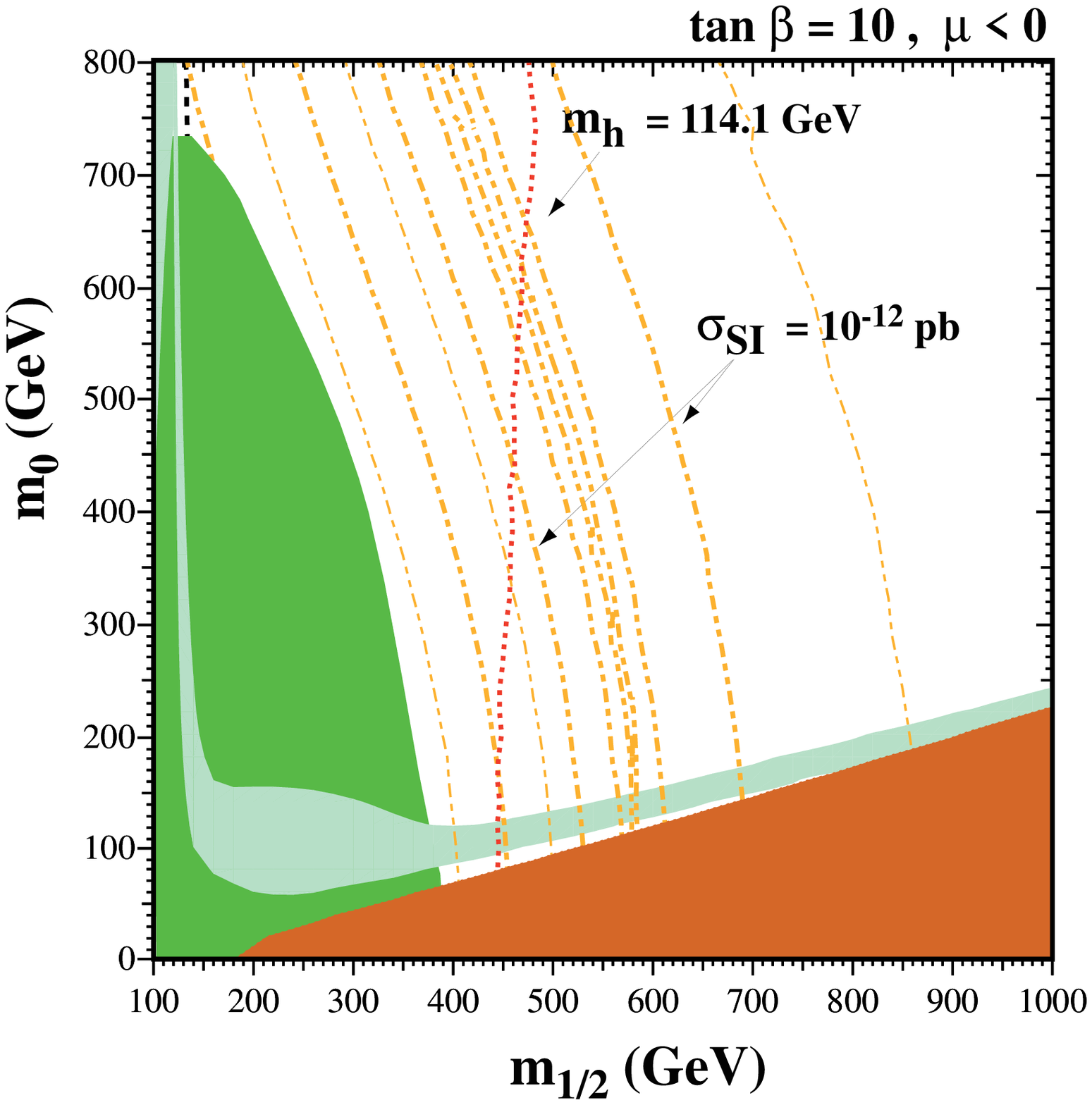,height=3.5in}
\hspace*{-0.17in}
\epsfig{file=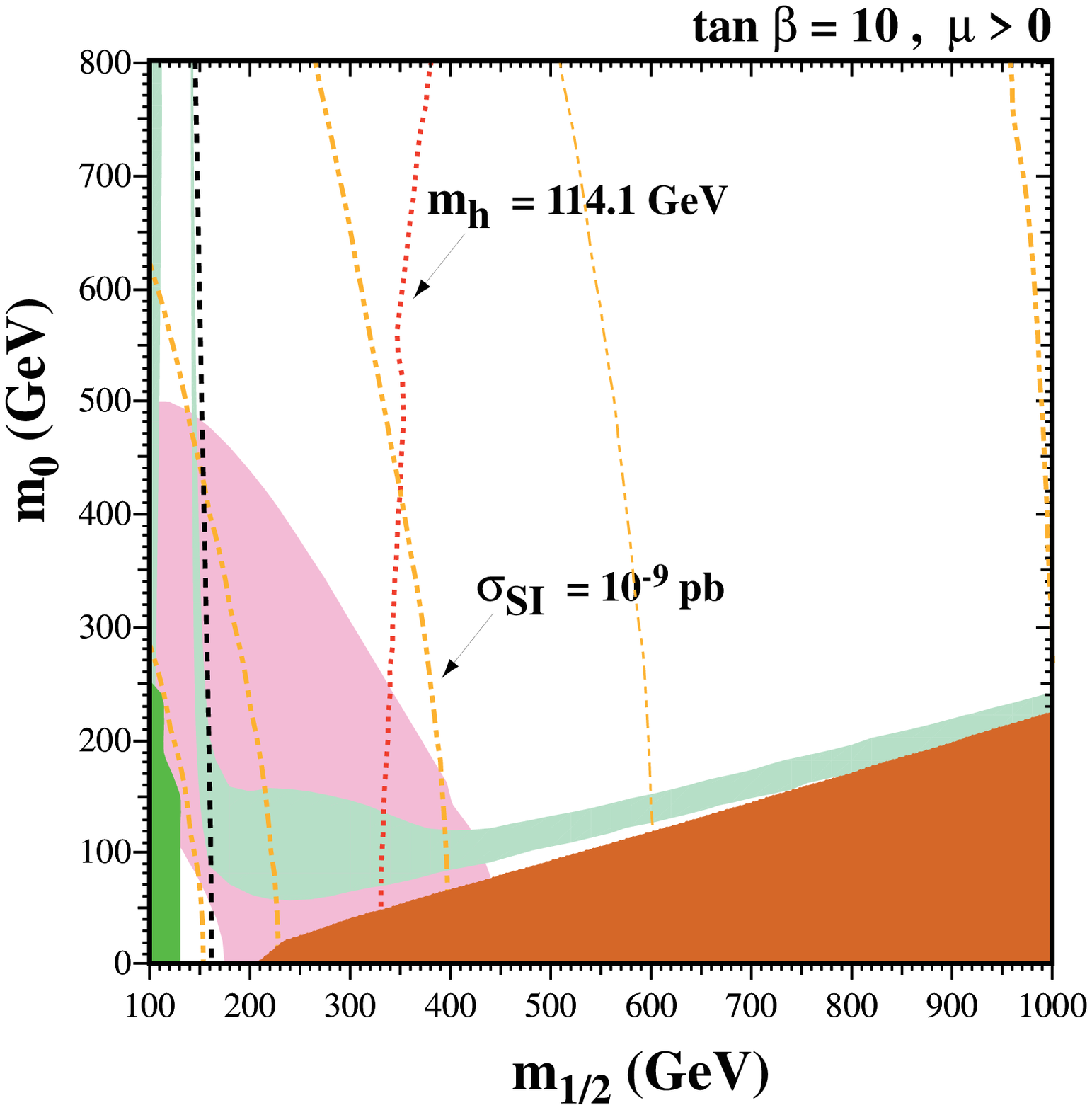,height=3.5in} \hfill
\end{minipage}
\begin{minipage}{8in}
\epsfig{file=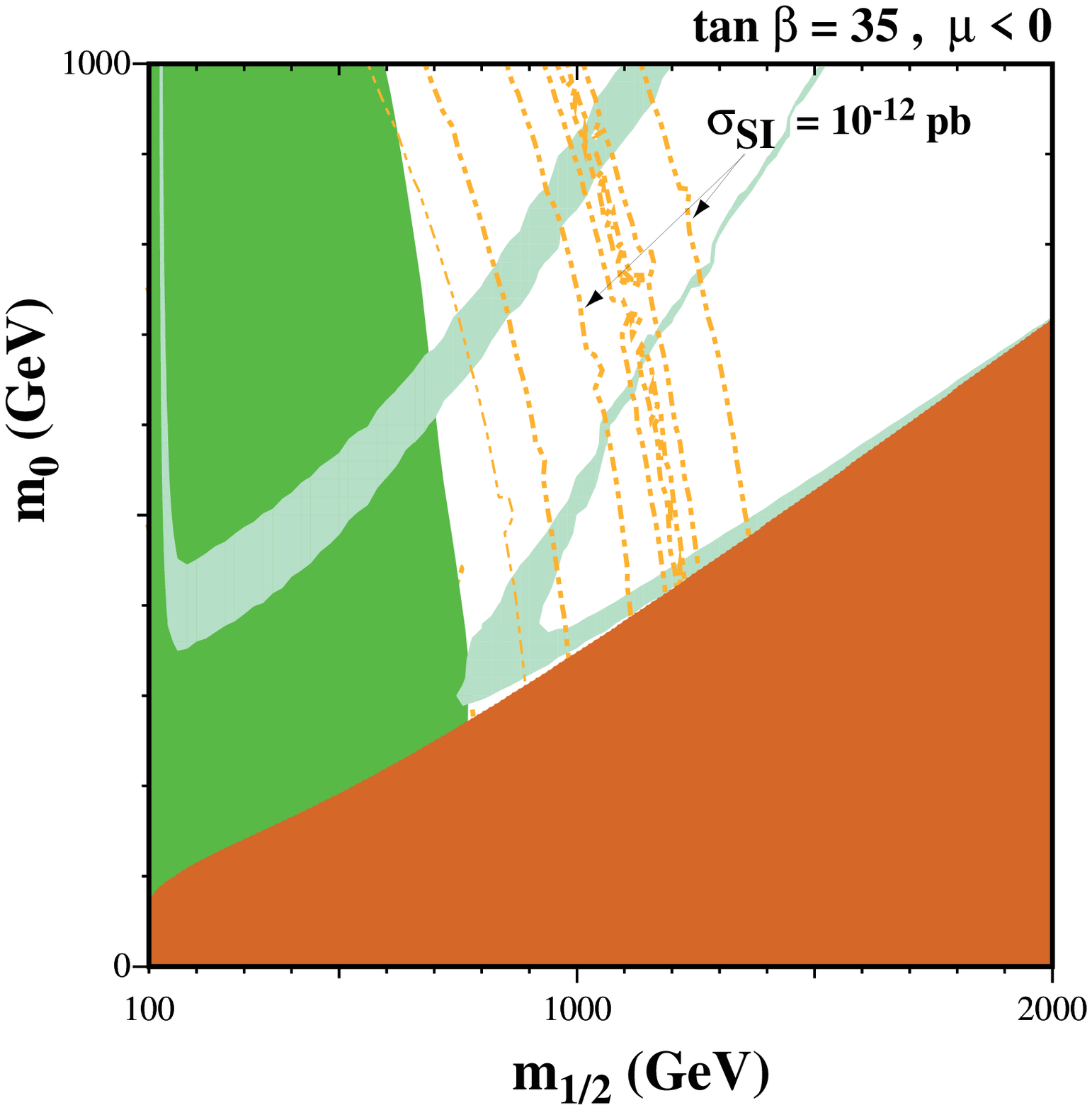,height=3.5in}
\hspace*{-0.2in}
\epsfig{file=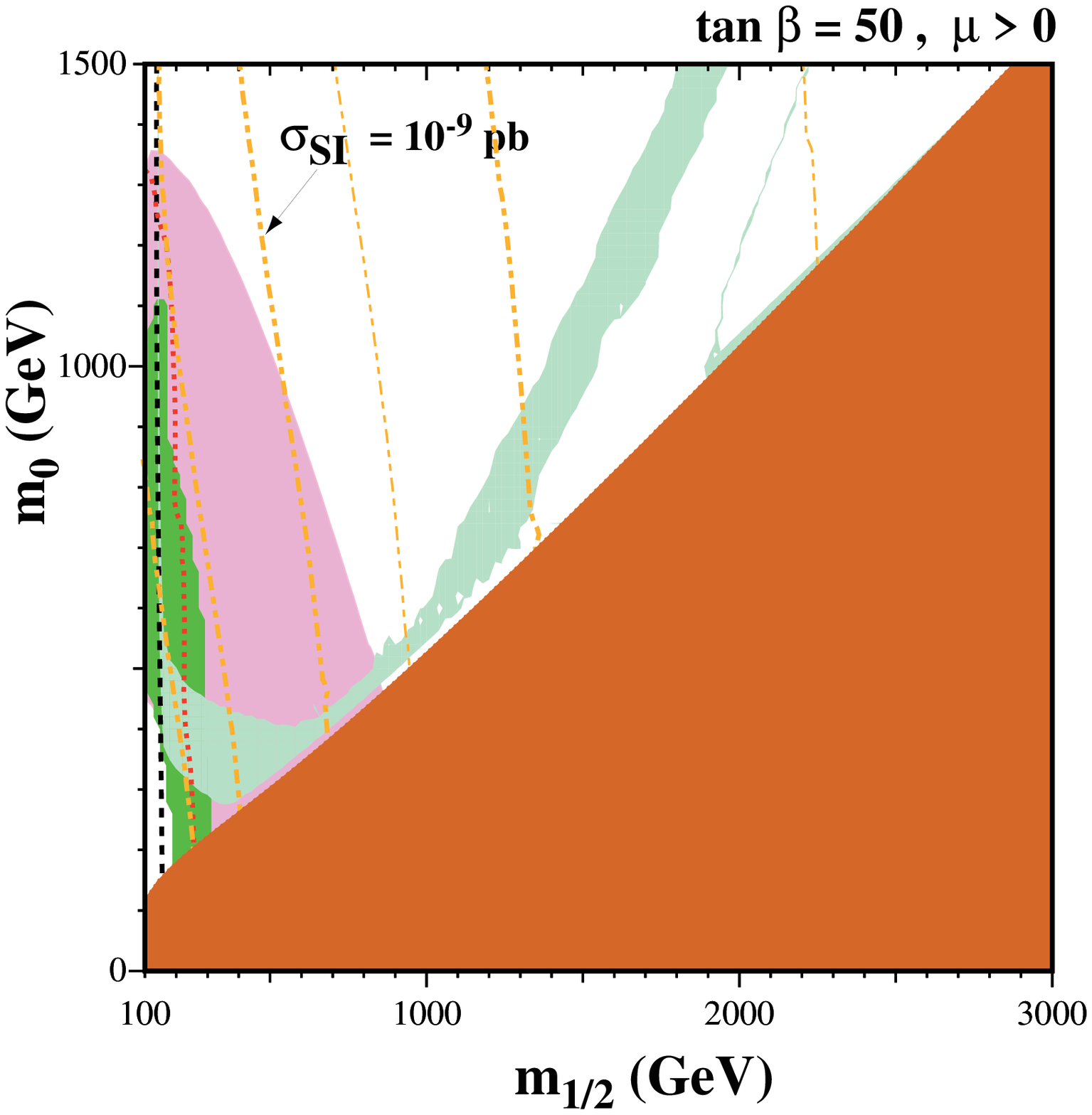,height=3.5in} \hfill
\end{minipage}
\caption{\label{fig:sicontours}
{\it Spin-independent cross sections in the $(m_{1/2}, m_0)$ planes for
(a) $\tan \beta = 10, \mu < 0$,  (b) $\tan \beta = 10, \mu > 0$, (c)
$\tan \beta = 35, \mu < 0$ and (d)  $\tan \beta = 50, \mu > 0$, assuming
$A_0 = 0, m_t = 175$~GeV and $m_b(m_b)^{\overline {MS}}_{SM} = 4.25$~GeV
\cite{EFGOSi}.   The double dot-dashed (orange) curves are contours of the
spin-independent cross section, differing by factors of 10 (bolder) and
interpolating factors of 3 (finer - when shown). For example, in (b), the
curves to the right of the one marked $10^{-9}$ pb correspond to $3
\times  10^{-10}$~pb and $10^{-10}$~pb. The near-vertical lines are the
LEP limits
$m_{\chi^\pm} = 103.5$~GeV (dashed and black)~\cite{LEPsusy}, $m_h =
114.1$~GeV (dotted and red)~\cite{LEPHiggs}. In the dark (brick red)
shaded regions, the LSP is the charged ${\tilde \tau}_1$, so this region
is excluded. The light (turquoise) shaded areas are the cosmologically
preferred regions with
\protect\mbox{$0.1\leq\ohsq\leq 0.3$}~\cite{EFGOSi}. The medium (dark
green) shaded regions that are most prominent in panels (a) and (c) are
excluded by $b \to s \gamma$~\cite{bsg}. The sloping shaded (pink) regions
in panels (b) and (d) delineate the $\pm 2 - \sigma$ ranges of $g_\mu -
2$~\cite{ENO}.
}}
\end{figure}

The code we use to calculate the elastic dark matter scattering cross
sections $\sigma_{SI,SD}$ was documented in~\cite{EFlO1,EFlO2}, together
with the ranges of values of the hadronic matrix elements that we use. The
cross sections for protons and neutrons are similar within the quoted
uncertainties in these matrix elements. Codes are
available~\cite{Neutdriver} that include additional contributions to the
scattering matrix elements, but a recent comparison~\cite{EFFMO} shows
that the improvements are not essential in the CMSSM parameter space that
we explore here. Fig.~\ref{fig:sicontours} displays contours of the
spin-independent cross section for the elastic scattering of the LSP
$\chi$ on protons in the $m_{1/2}, m_0$ planes for (a) $\tan \beta = 10,
\mu < 0$, (b) $\tan \beta = 10, \mu > 0$, (c) $\tan \beta = 35, \mu < 0$,
and (d) $\tan \beta = 50, \mu > 0$. The latter are
close to the largest values of $\tan \beta$ for which we find generic
solutions to the electroweak symmetry-breaking conditions for $\mu < 0$
and $> 0$, respectively~\cite{EFGOSi}. The double dot-dashed (orange)
lines are contours of the spin-independent cross section, and we have
indicated the contours
$\sigma_{SI} = 10^{-9}$~pb in panels (a, d) and $\sigma_{SI} =
10^{-12}$~pb in panels (b, c). The other bolder contours are
for cross sections differing by factors of 10, and the finer 
contours for cross sections differing by interpolating factors of 3
(in order to ensure clarity, not all of the interpolating contours are 
displayed).

These cross-section contours are combined in Fig.~\ref{fig:sicontours}
with other information on the CMSSM parameter space. The lower right-hand
corners of the panels are excluded because there the LSP is the lighter
$\tilde \tau_1$. The light (turquoise) shaded regions are those with $0.1 
< \Omega_\chi h^2 < 0.3$~\cite{omegah2}. The `bulk' regions are clearly
visible in panels (a,b) and (d), and coannihilation `tails' in all
panels~\cite{EFGOSi}. For our default choices $A = 0$, $m_t(pole) =
175$~GeV and $m_b(m_b)^{\overline {MS}}_{SM} = 4.25$~GeV, the
`focus-point' regions~\cite{focus} are at larger values of $m_0$ than are
shown in any of the panels. Rapid-annihilation `funnels' are visible in
panels (c) and (d): that in the former panel bisects the `bulk' region.
The near-vertical dashed (black) lines at small $m_{1/2}$ are the
chargino-mass contours
$m_{\chi^\pm} = 103.5$~GeV~\cite{LEPsusy}, and the near-vertical dotted
(red) lines at larger $m_{1/2}$ are the contours $m_h =
114.1$~GeV~\cite{LEPHiggs}, as calculated using {\tt 
FeynHiggs}~\cite{FeynHiggs}. The large medium (green)  shaded regions in
panels (a) and (c) are those excluded by $b \to s \gamma$~\cite{bsg}:
smaller excluded regions are also visible in panels (b) and (d)  at small
$m_{1/2}$. Finally, the sloping shaded (pink) regions in panels (b) and
(d) delineate the $\pm 2 - \sigma$ ranges of $g_\mu - 2$~\cite{ENO}, which
are absent for the $\mu < 0$ panels (a) and (c).

The LEP lower limits on $m_h$ and $m_{\chi^\pm}$, as well as the
experimental measurement of $b \to s \gamma$ for $\mu < 0$, tend to bound
the cross sections from above, as we discuss later in more
detail. Generally speaking, the spin-independent cross section is 
relatively large in the `bulk' region, but falls off in the `tail' and
`funnel' regions. In the focus-point regions, the spin-independent cross
section is relatively independent of $m_\chi$ and for $\tan \beta = 10$,
takes a value between $10^{-9}$ and $10^{-8}$ pb \cite{focus,EFFMO}.  
Also, we note also that there is a strong cancellation in the
spin-independent cross section when
$\mu < 0$~\cite{EFlO1,EFlO2}, as seen along strips in panels (a, c)  of
Fig.~\ref{fig:sicontours} where
$m_{1/2} \sim 500, 1100$~GeV, respectively. In the cancellation region,
the cross section drops lower than $10^{-14}$ pb. All these possibilities
for suppressed spin-independent cross sections are disfavoured by the
data on
$g_\mu - 2$~\cite{g-2,susyg-2,ENO}, which favour values of $m_{1/2}$ and
$m_0$ that are not very large, as well as $\mu > 0$, as seen in panels
(b, d) of Fig.~\ref{fig:sicontours}. Thus $g_\mu - 2$ tends to provide a
lower bound on the spin-independent cross section. 

\begin{figure}
\vspace*{-0.75in}
\begin{minipage}{8in}
\epsfig{file=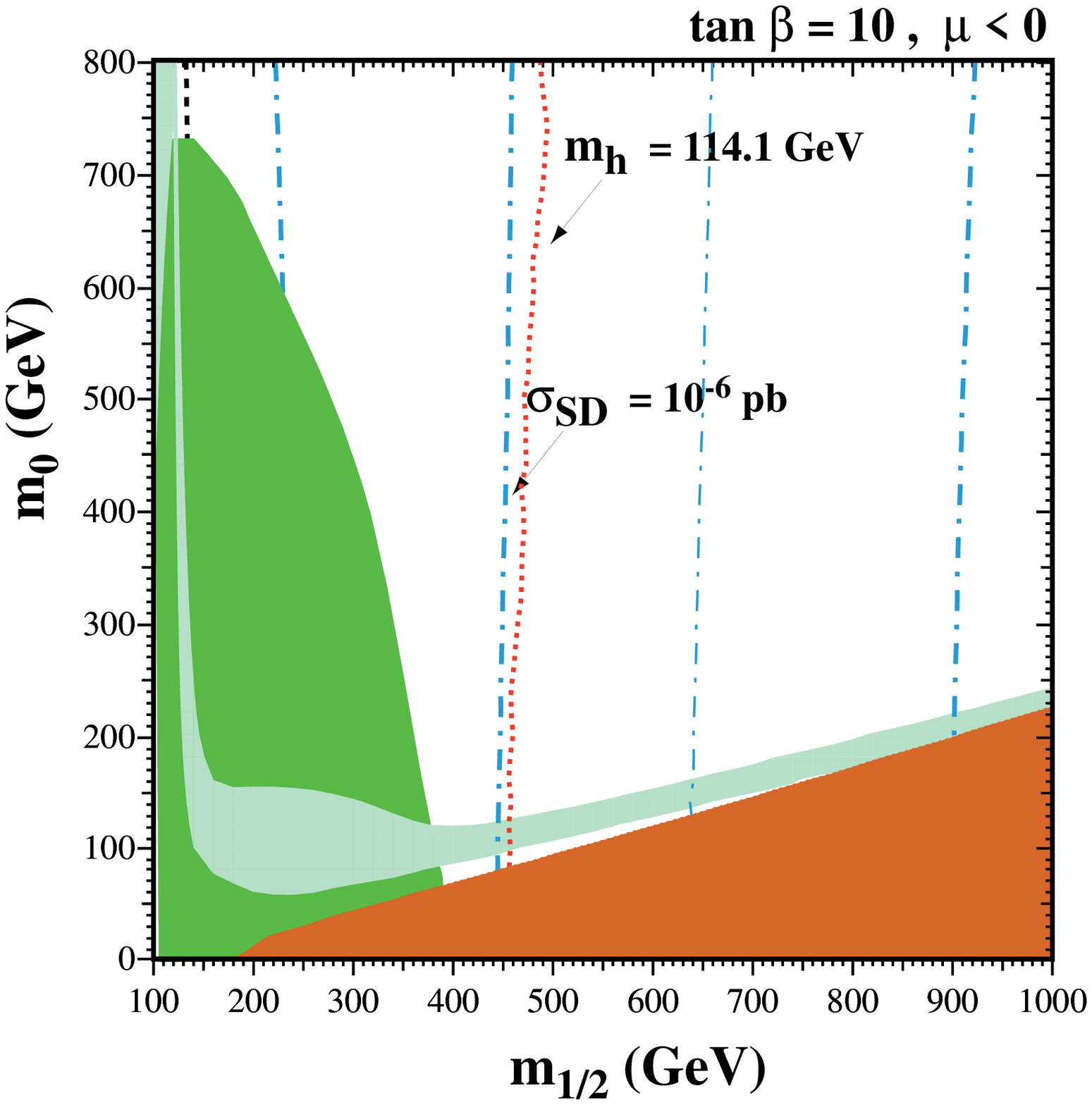,height=3.5in}
\hspace*{-0.17in}
\epsfig{file=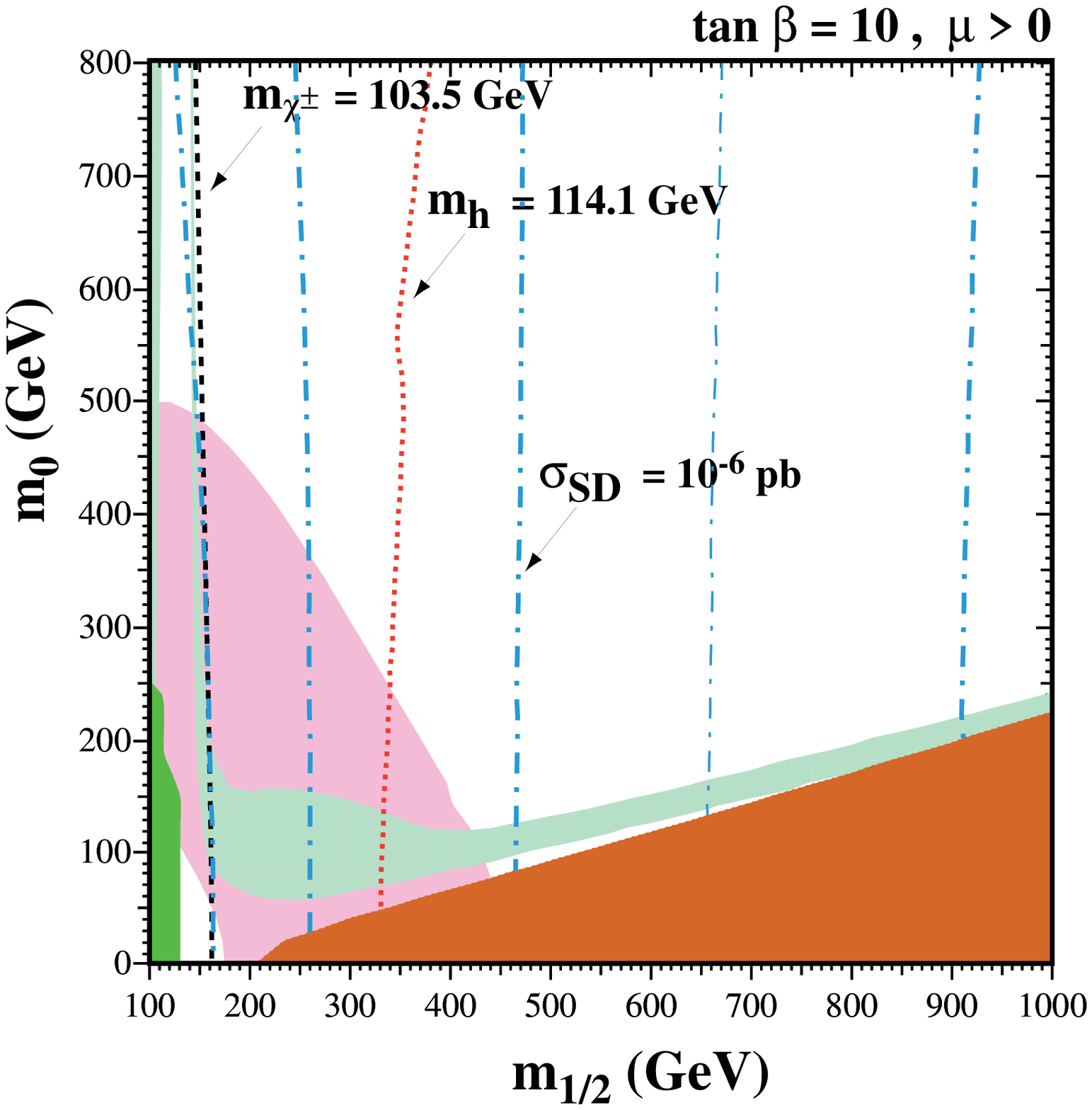,height=3.5in} \hfill
\end{minipage}
\begin{minipage}{8in}
\epsfig{file=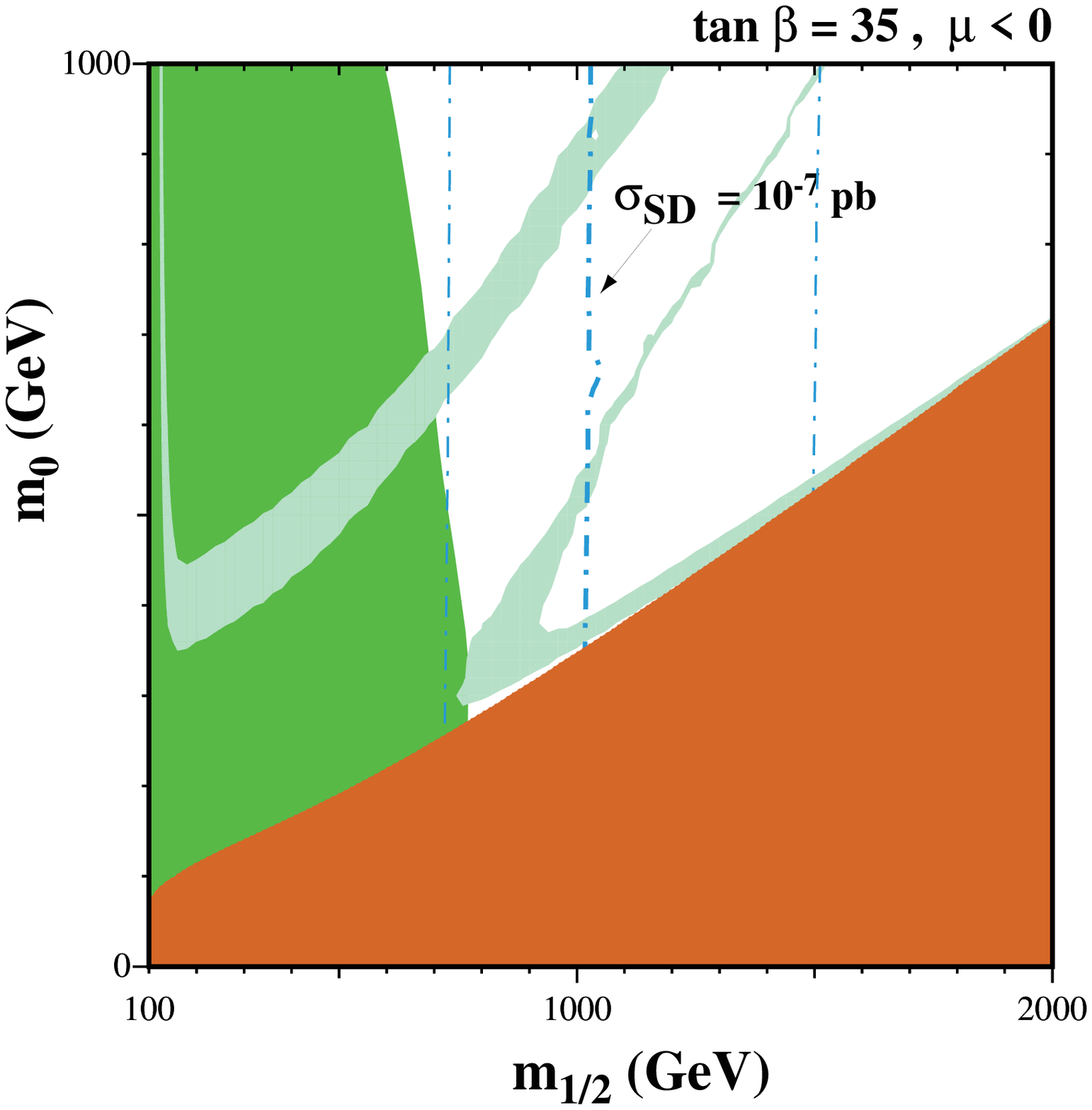,height=3.5in}
\hspace*{-0.2in}
\epsfig{file=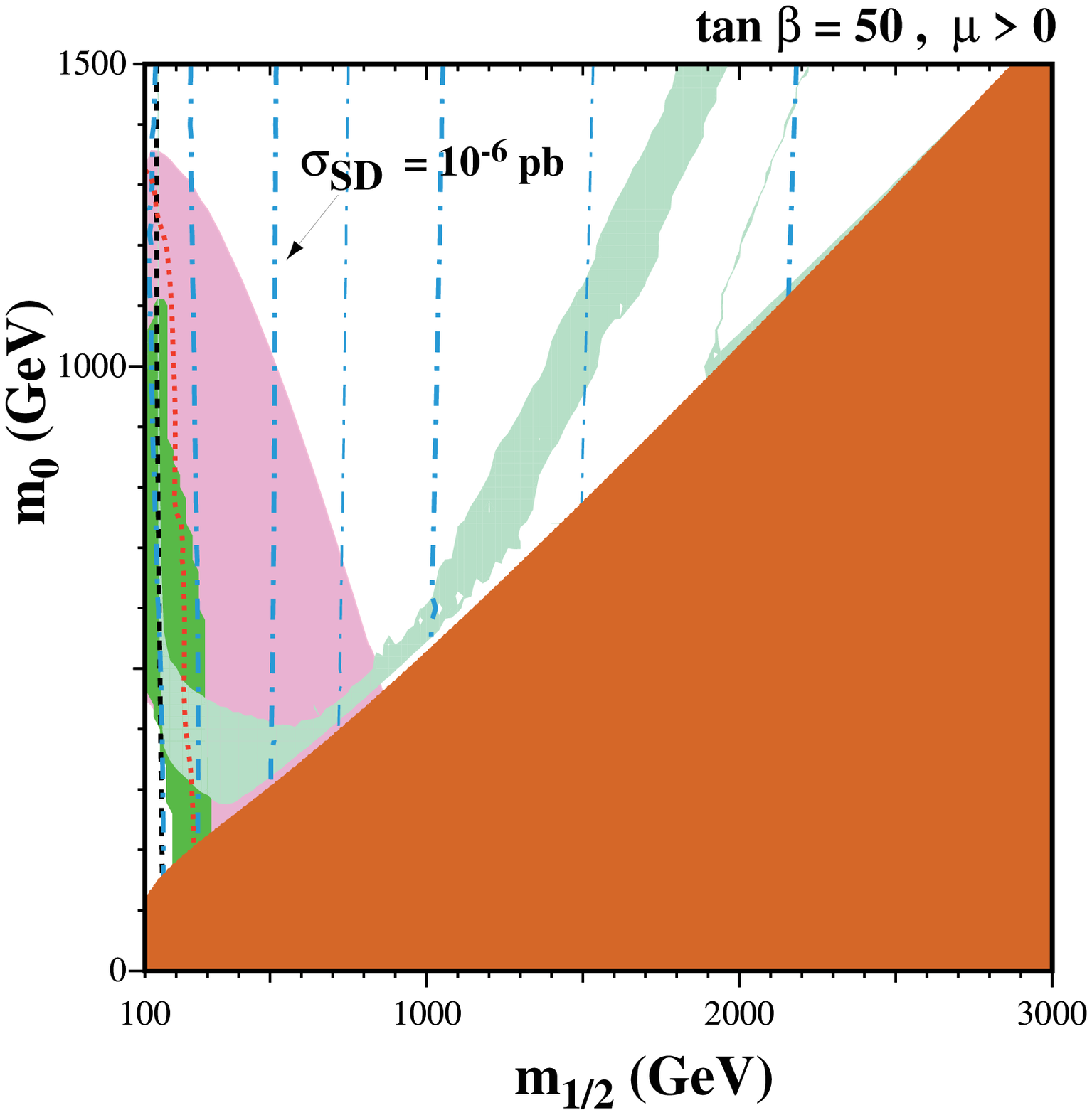,height=3.5in} \hfill
\end{minipage}
\caption{\label{fig:sdcontours}
{\it Spin-dependent cross sections in the $(m_{1/2}, m_0)$ planes for (a)
$\tan \beta = 10, \mu < 0$, (b) $\tan \beta = 10, \mu > 0$, (c) $\tan
\beta = 35, \mu < 0$ and (d)  $\tan \beta = 50, \mu > 0$, assuming $A_0 =
0, m_t = 175$~GeV and $m_b(m_b)^{\overline {MS}}_{SM} =
4.25$~GeV~\cite{EFGOSi}.  The dot-dashed (blue) lines are contours of the
spin-dependent cross section, differing by factors of 10 (bolder) and
interpolating factors of 3 (finer -  when shown). The near-vertical dashed
lines are the LEP limits $m_{\chi^\pm} = 103.5$~GeV
(black)~\cite{LEPsusy},
$m_h = 114.1$~GeV (red)~\cite{LEPHiggs}. In the dark (brick red) shaded
regions, the LSP is the charged ${\tilde \tau}_1$, so this region is
excluded. The light (turquoise) shaded areas are the cosmologically
preferred regions with \protect\mbox{$0.1\leq\ohsq\leq
0.3$}~\cite{omegah2}.  The medium (dark green) shaded regions are
excluded by $b \to s \gamma$~\cite{bsg}.}}
\end{figure}

Fig.~\ref{fig:sdcontours} displays contours of the spin-dependent cross
section in the $m_{1/2}, m_0$ planes for (a) $\tan \beta = 10, \mu < 0$,
(b) $\tan \beta = 10, \mu > 0$, (c) $\tan \beta = 35, \mu < 0$, and (d)
$\tan \beta = 50, \mu > 0$. The dot-dashed (blue) lines are those of the
spin-dependent cross section, and the other notation is as in
Fig.~\ref{fig:sicontours}. The bolder lines are contours differing by
factors of 10 from the indicated ones, and the finer lines, when shown, 
differ by interpolating factors of 3. We note again that the cross
section is generically larger in the `bulk' region and smaller in the
coannihilation `tail' and rapid-annihilation `funnel' regions. In the
focus-point regions, the spin-dependent cross-section is also relatively
constant and for
$\tan \beta = 10$ takes values between $10^{-5}$ and $10^{-4}$ pb
\cite{focus,EFFMO}. Unlike the spin-independent case, there are no
cancellations in the spin-dependent cross section.

\begin{figure}
\vspace*{-0.75in}
\begin{minipage}{6in}
\begin{center}
\hspace*{-0.4in}
\epsfig{file=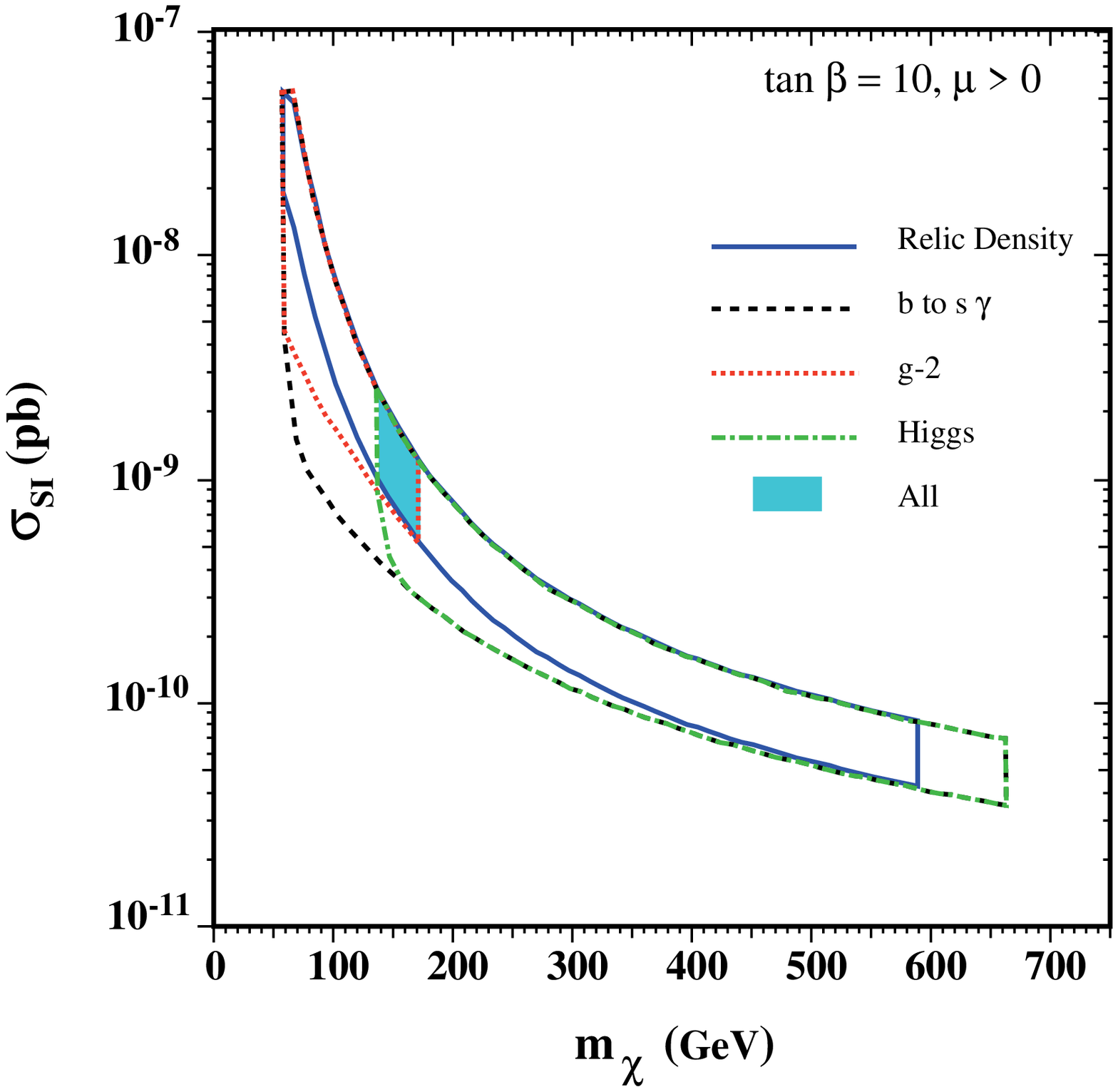,height=2.5in}
\hspace*{0.1in}
\epsfig{file=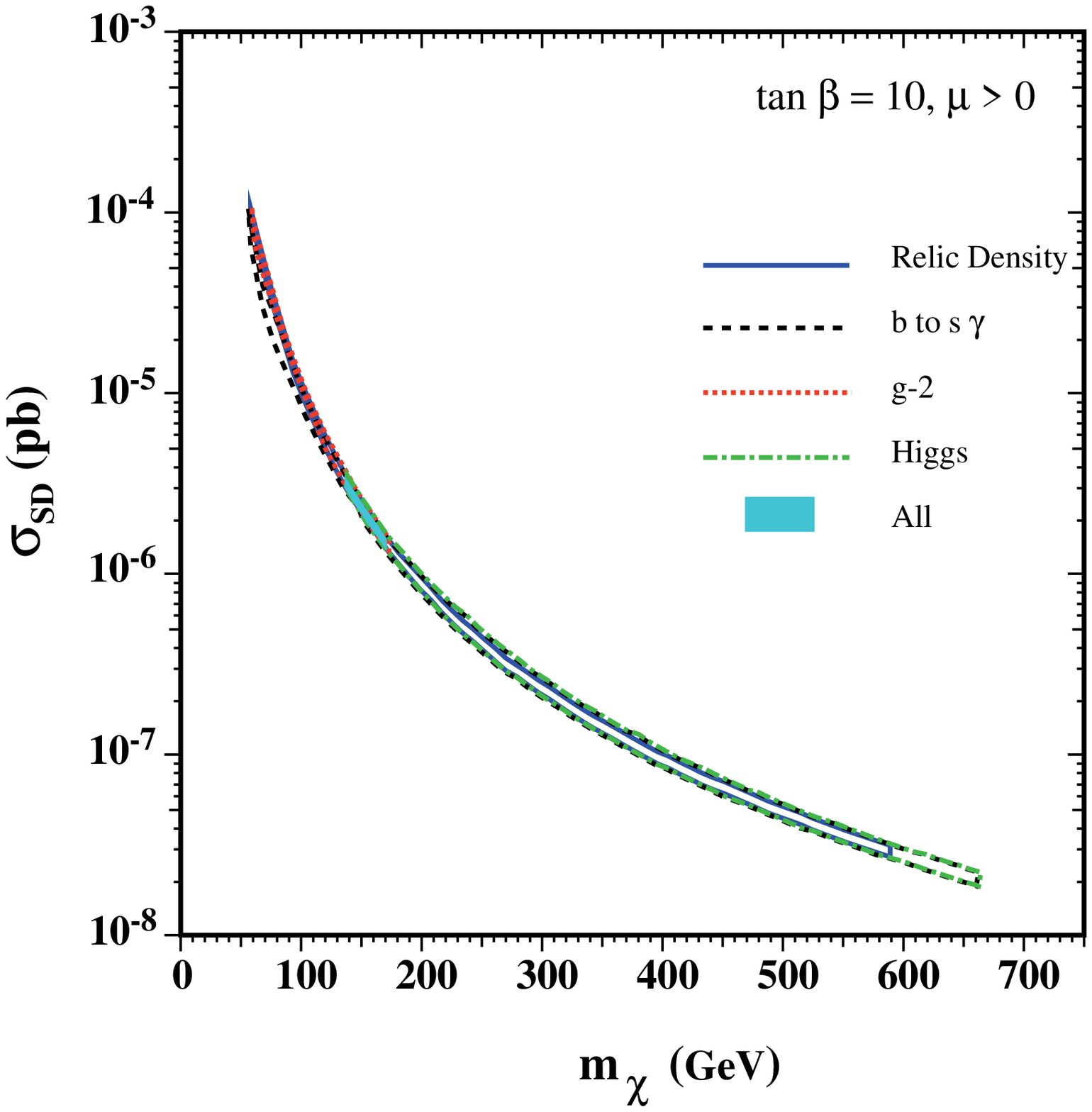,height=2.5in} \hfill
\end{center}
\end{minipage}
\hspace*{-.50in}
\begin{minipage}{6in}
\begin{center}
\hskip -1.20in
\epsfig{file=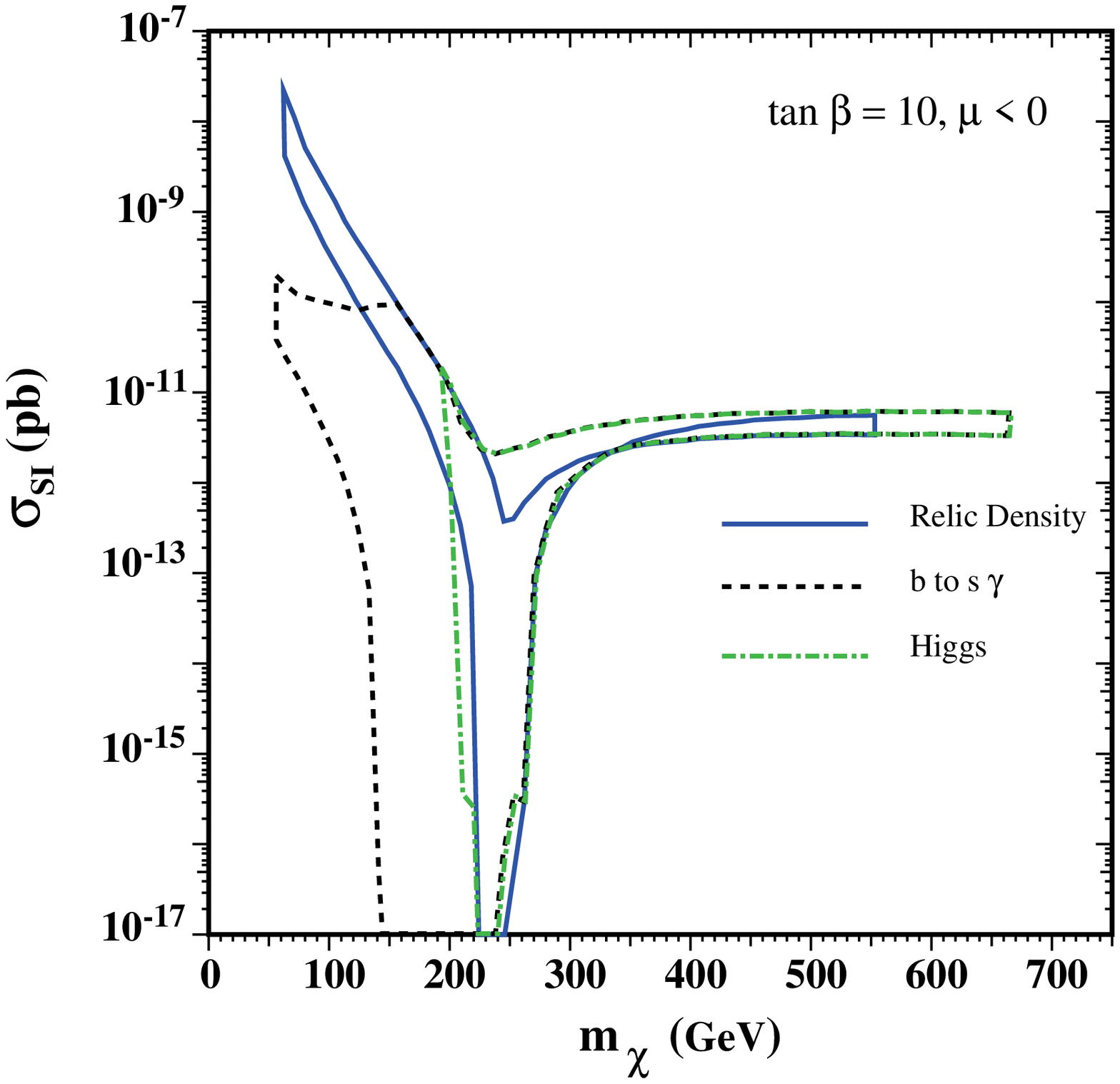,height=2.5in}
\hspace*{0.1in}
\epsfig{file=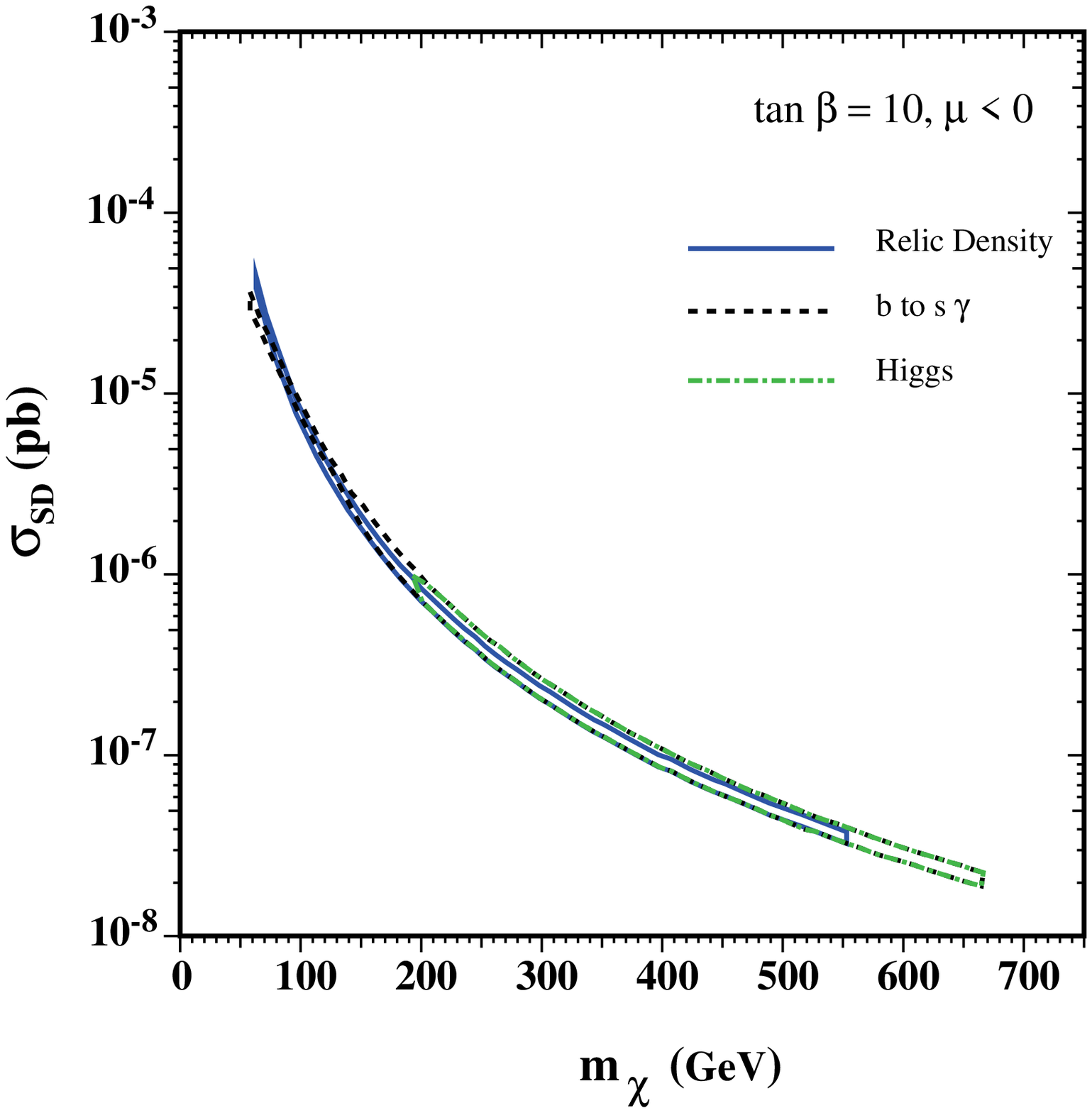,height=2.5in} \hfill
\end{center}
\end{minipage}
\begin{minipage}{6in}
\begin{center}
\epsfig{file=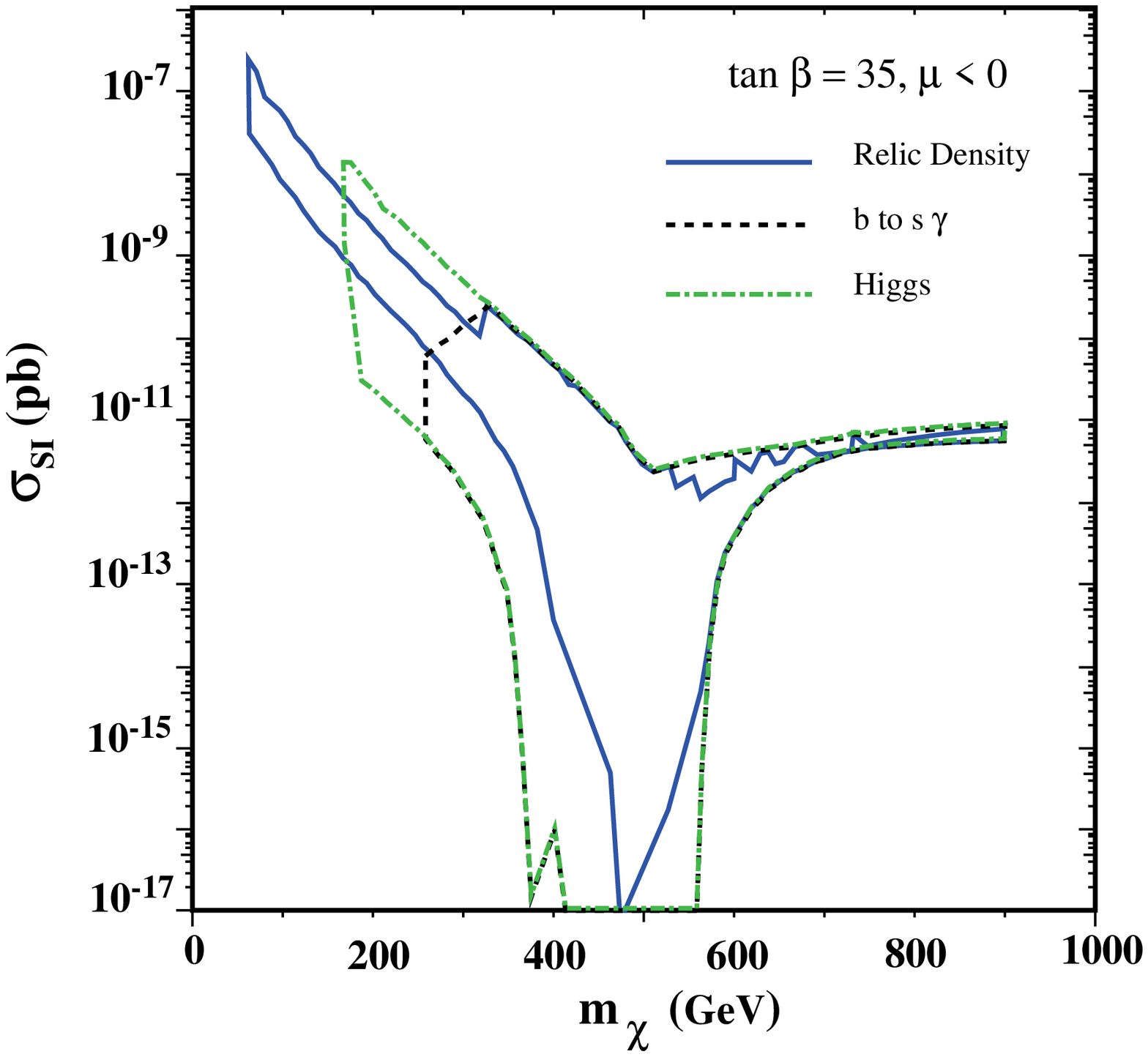,height=2.5in}
\epsfig{file=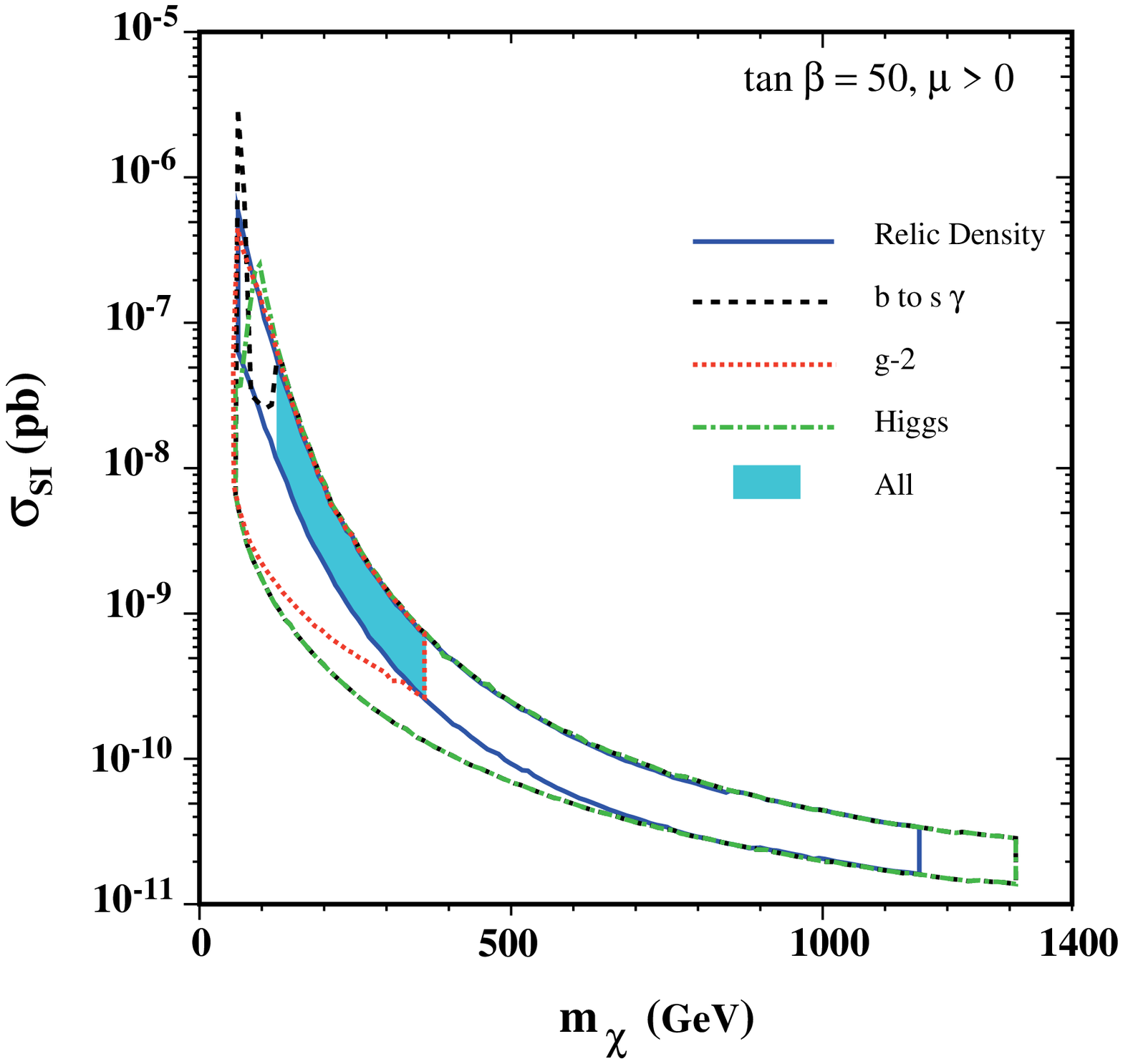,height=2.5in} \hfill
\end{center}
\end{minipage}
\caption{\label{fig:decimation}
{\it Allowed ranges of the cross sections for $\tan \beta = 10$ and (a, b)
$\mu > 0$, (c, d) $\mu < 0$, for (a, c) spin-independent and (b, d)
spin-dependent elastic scattering. Panel (e) shows the spin-independent 
cross section for $\tan \beta = 35$ and $\mu < 0$, and panel (f) the
spin-independent   
cross section for $\tan \beta = 50$ and $\mu > 0$. The solid (blue) lines 
indicate the
relic density constraint~\cite{omegah2}, the dashed (black) lines the $b
\to s \gamma$ constraint~\cite{bsg}, the dot-dashed (green) lines the
$m_h$ constraint~\cite{LEPHiggs}, and the dotted (red) lines the $g_\mu -
2$ constraint~\cite{ENO}. The shaded (pale blue)  region is allowed by all
the constraints.
}}
\end{figure}

Fig.~\ref{fig:decimation} illustrates the effect on the cross sections of
each of the principal phenomenological constraints, in the particular
cases $\tan \beta = 10$ and (a, b) $\mu > 0$, (c, d) $\mu < 0$, 
(e) $\tan \beta = 35, \mu < 0$ and (f) $\tan \beta = 50, \mu > 0$. The solid
(blue) lines mark the bounds on the cross sections allowed by the
relic-density constraint $0.1 < \Omega_\chi h^2 < 0.3$
alone~\cite{omegah2}. 
For any given value of
$m_{1/2}$, only a restricted range of $m_0$ is allowed. Therefore, only a
limited range of $m_0$, and hence only a limited range for the cross
section, is allowed for any given value of
$m_\chi$. The thicknesses of the allowed regions are due in part to the 
assumed
uncertainties in the nuclear inputs.  These have been discussed at
legnth in \cite{EFlO2,EFlO1} and we refer the reader there for details.
In  the case (e) of $\tan \beta = 35, \mu < 0$ and (f) of $\tan \beta =
50, \mu > 0$, two or three different narrow  ranges of $m_0$ may be
allowed for the same value of
$m_{1/2}$, but they  have quite similar cross sections, as seen already
in  Figs.~\ref{fig:sicontours}(c,d) and \ref{fig:sdcontours}(c,d). On 
the other hand, a broad range of $m_\chi$ is allowed, when one takes into
account the coannihilation `tail' region at each $\tan 
\beta$ and the rapid-annihilation `funnel' regions for $\tan
\beta = 35, 50$~\cite{EFGOSi}~\footnote{We do not show predictions in the 
`focus-point' region~\cite{focus}.}. The dashed (black) lines in
Fig.~\ref{fig:decimation} display the range allowed by the $b \to s
\gamma$ constraint~\cite{bsg} alone, which is more important for $\mu < 
0$. In
this case, a broader range of $m_0$ and hence the spin-independent cross
section is possible for any given value of $m_\chi$. The impact of the
constraint due to $m_h$ is shown by the dot-dashed (green) lines in
Fig.~\ref{fig:decimation}. We implement this constraint by requiring that
$m_h > 114.1$~GeV when calculated using the {\tt FeynHiggs}
code~\cite{FeynHiggs}. Comparing with the previous constraints, we see
that a region at low $m_\chi$ is excluded by $m_h$, strengthening
significantly the previous {\it upper} limit on the spin-independent cross
section. Finally, the dotted (red)  lines in Fig.~\ref{fig:decimation}
show the impact of the $g_\mu - 2$ constraint~\cite{ENO}. This imposes an
upper bound on
$m_{1/2}$ and hence $m_\chi$, and correspondingly a {\it lower} limit on
the spin-independent cross section.

We emphasize again the important impacts of the updated LEP limits on the
chargino and (particularly) Higgs masses. Significantly smaller LSP masses
and correspondingly larger cross sections could be found if one used
earlier, weaker LEP limits.

The shaded (pale blue) regions in panels (a,b,f) of 
Fig.~\ref{fig:decimation} show the ranges of
$m_\chi$ and the cross sections that 
survive all the
phenomenological constraints. We find for $\tan \beta = 10$, 
\begin{equation}
135~{\rm GeV} \la m_\chi \la 180~{\rm GeV} \; {\rm for} \; \mu > 0
\label{mchirange}
\end{equation}
and the lower limit is $m_\chi \ga 190$~GeV for $\mu < 0$. The upper
bound  in 
(\ref{mchirange}) is due to $g_\mu - 2$, and there is no such upper bound 
for $\mu < 0$, unless one interprets the LEP `hint' as a real Higgs 
signal~\cite{LEPHiggs}, and imposes $m_h < 117$~GeV, in which case one 
finds $m_\chi \la 370$~GeV. The ranges of cross sections corresponding to 
(\ref{mchirange}) are 
\begin{eqnarray}
5 \times 10^{-10}~{\rm pb} \la & \sigma_{SI} & \la 3 \times 10^{-9}~{\rm
pb}, 
\\
1 \times 10^{-6}~{\rm pb} \la & \sigma_{SD} & \la 4 \times 10^{-6}~{\rm
pb},
\label{xsecrangep}
\end{eqnarray}
for $\tan \beta = 10$ and $\mu > 0$, and we find
\begin{eqnarray}
\sigma_{SI} & \la & 2 \times 10^{-11}~{\rm pb}, \\
\sigma_{SD} & \la & 1 \times 10^{-6}~{\rm pb}, 
\label{xsecrangen}
\end{eqnarray}
for $\tan \beta = 10$ and $\mu < 0$. No lower limits for the
spin-independent cross section are possible with 
the constraints considered above, both  because the $g_\mu -
2$ constraint is inapplicable and must be discarded if this sign of $\mu$
is to be considered at all, and also  because of the cancellation in
$\sigma_{SI}$ that is visible in panels (a)  and (c) of
Fig.~\ref{fig:sicontours}. Even if we take the LEP `hint' 
of a signal for a Higgs boson, and impose the upper limit $m_h 
< 117$~GeV, because the bound on $m_\chi$ is past the cancellation region,
we find no useful lower bound for $\tan
\beta = 10$ and
$\mu <  0$. For the spin-dependent cross section,
a lower limit due to the relic density is determined by the endpoint of 
the coannihilation region,
namely $\sigma_{SD} \ga 2 \times 10^{-8}$ pb. A Higgs mass bound of 117
GeV in this case would impose $\sigma_{SD} \ga  10^{-7}$ pb.

\begin{figure}  
\vspace*{-0.75in}
\begin{minipage}{8in}
\epsfig{file=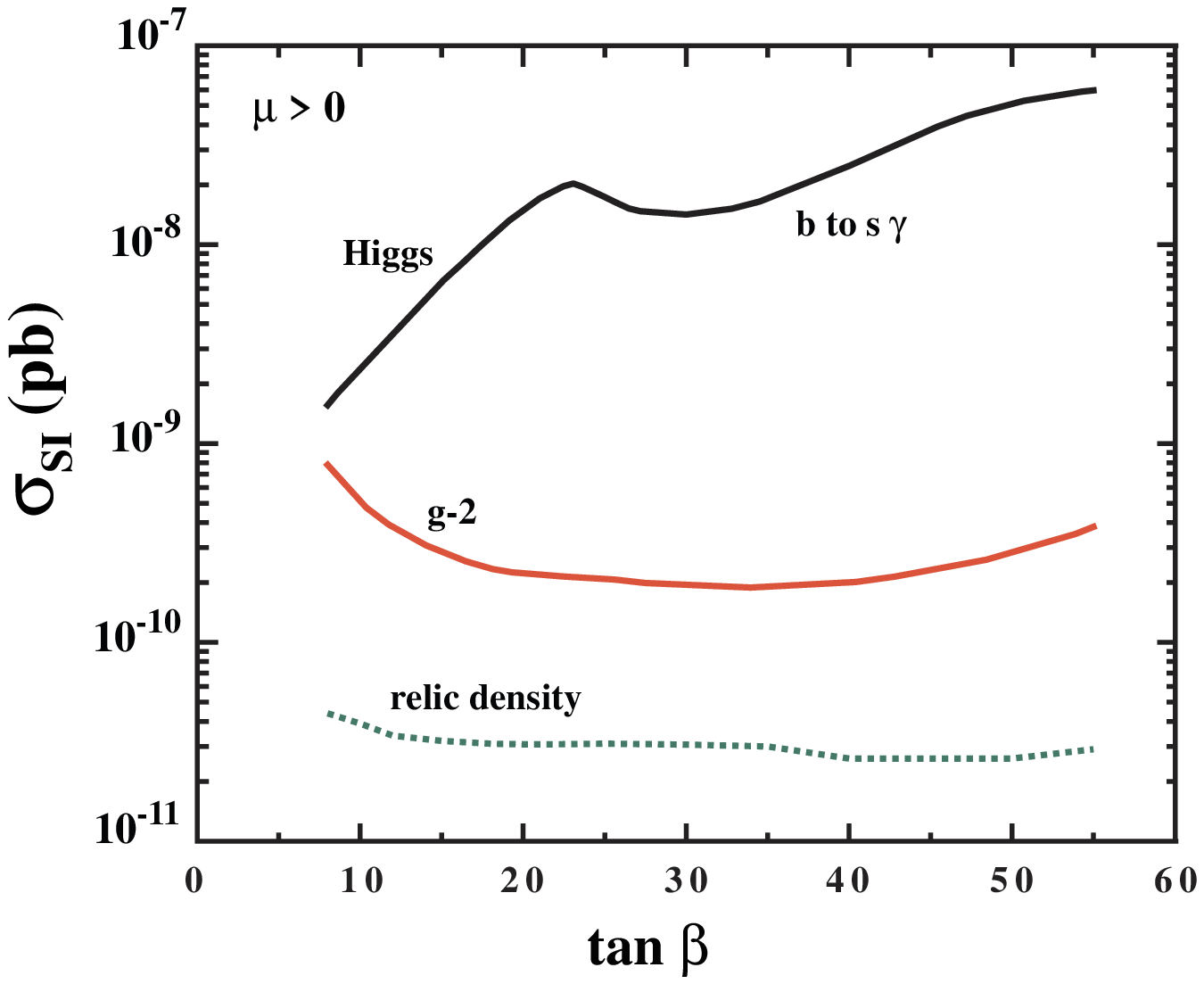,height=2.5in}
\hspace*{-0.17in}
\epsfig{file=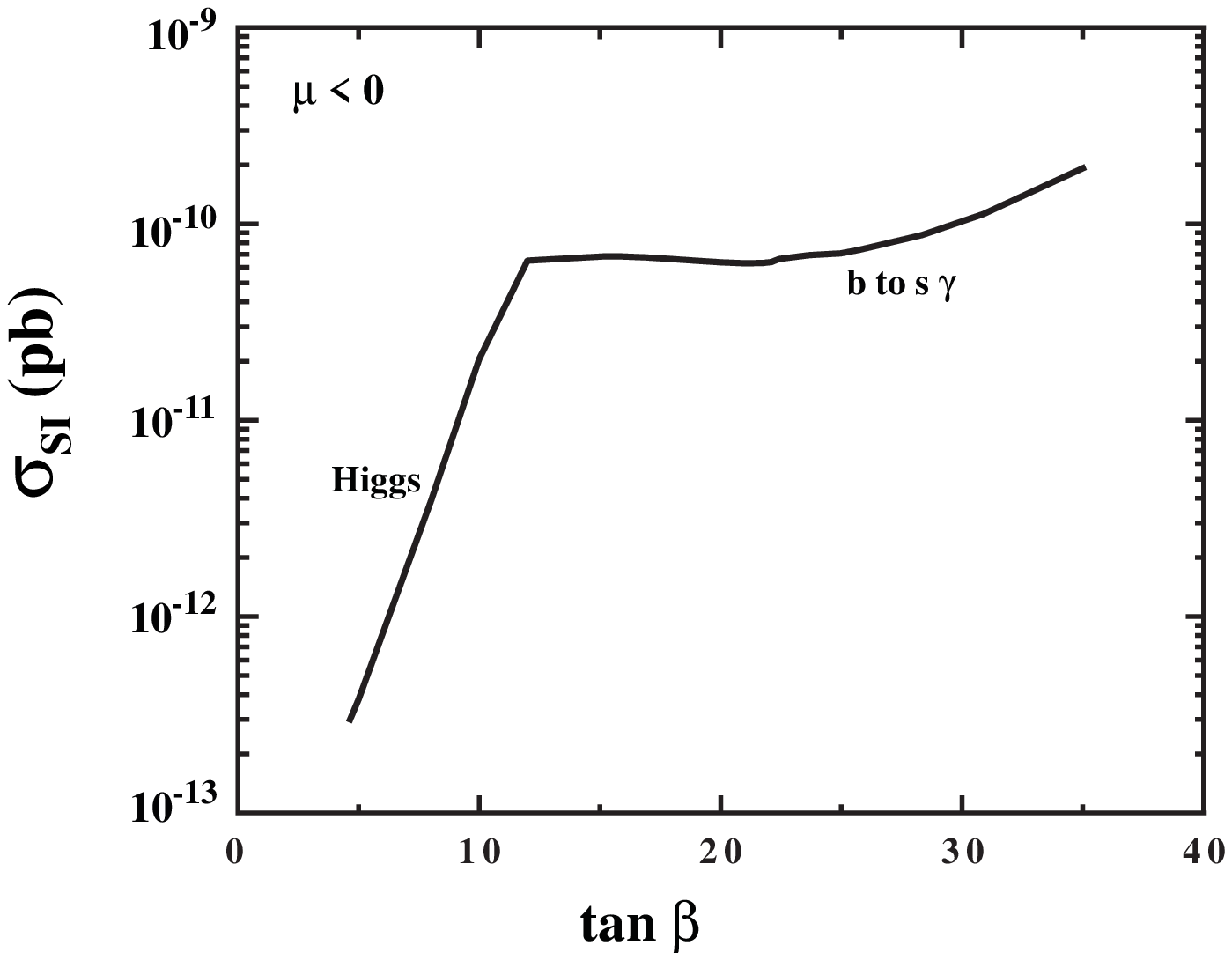,height=2.5in} \hfill
\end{minipage}   
\begin{minipage}{8in}
\epsfig{file=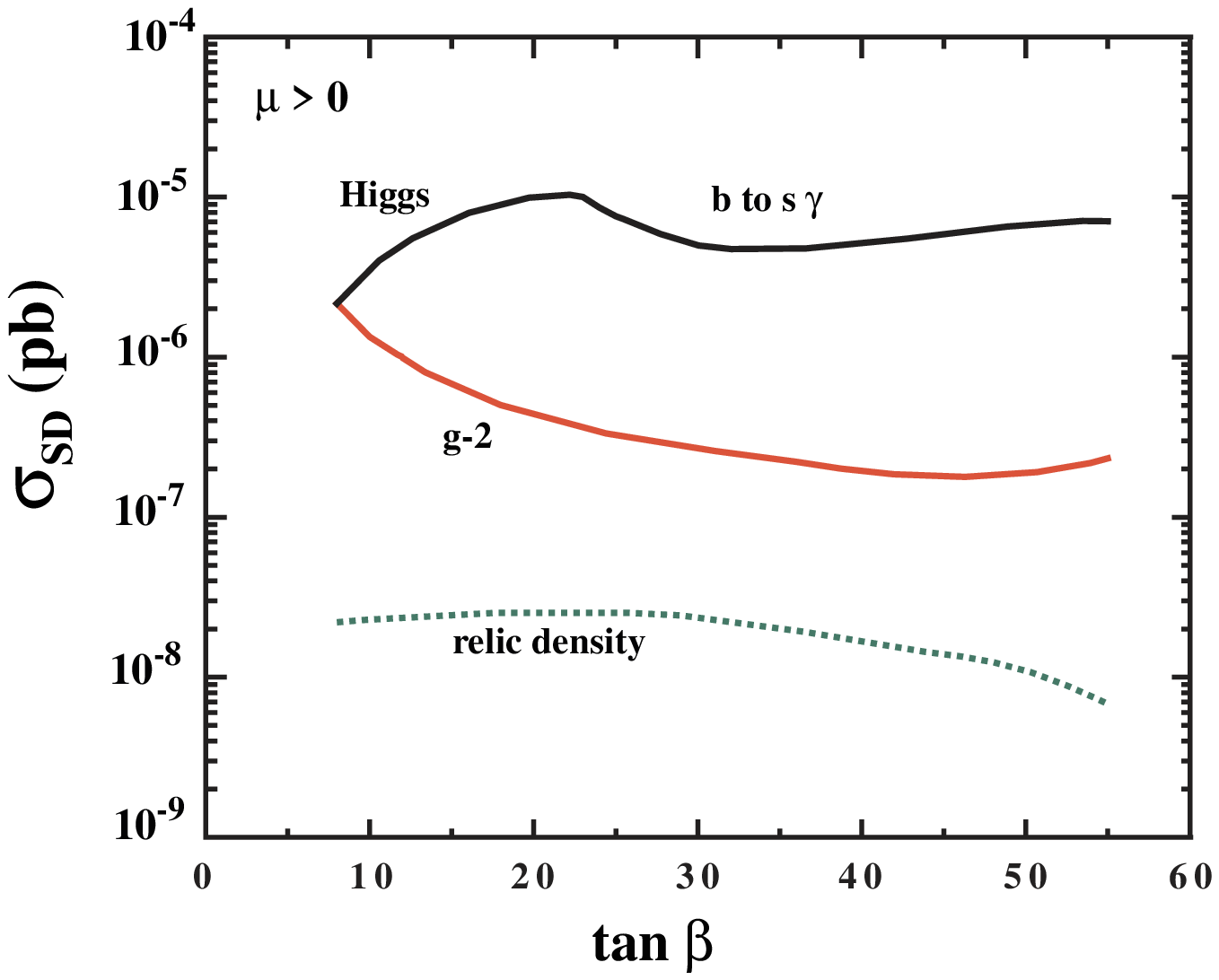,height=2.5in}
\hspace*{-0.2in}
\epsfig{file=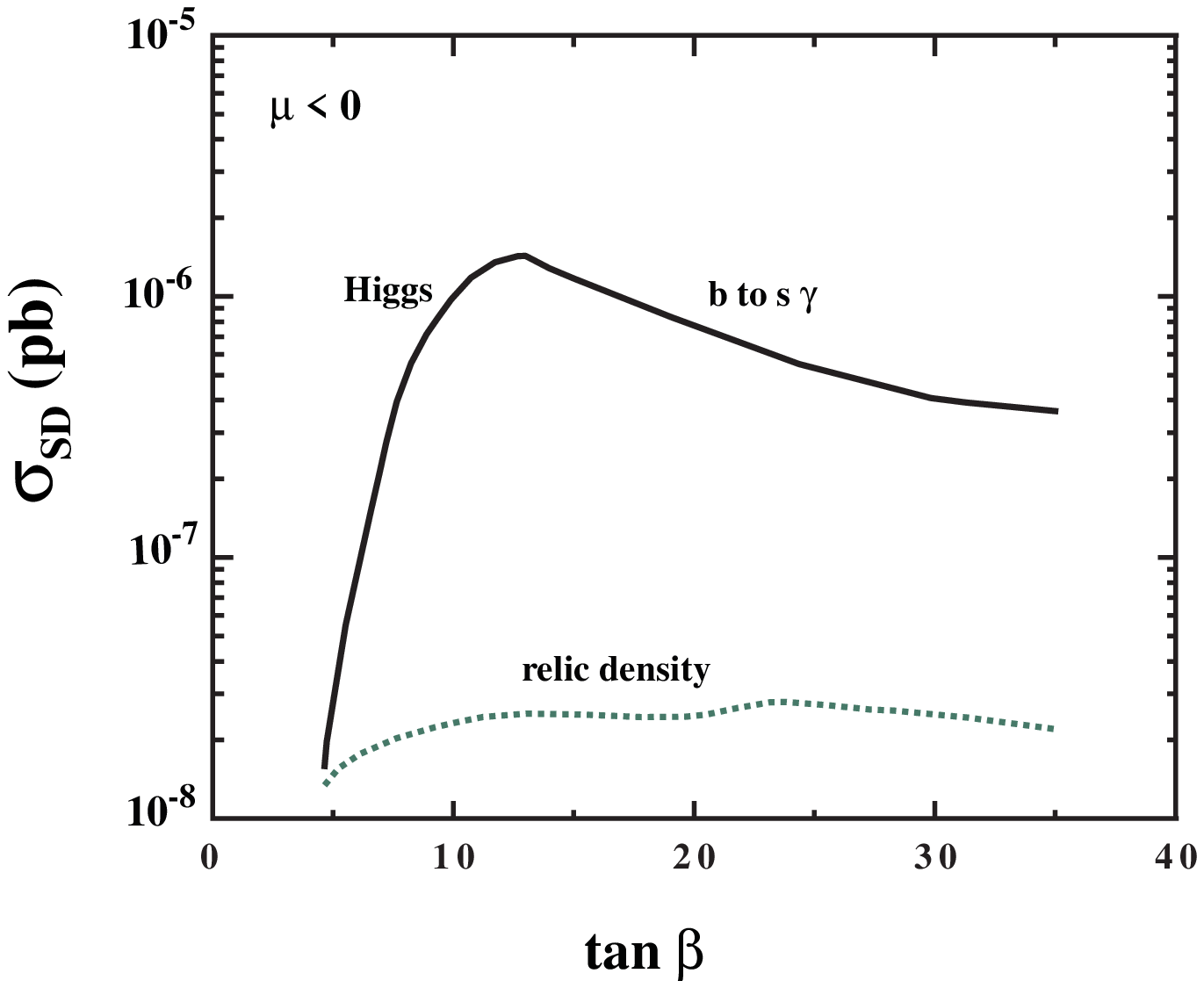,height=2.5in} \hfill
\end{minipage}
\caption{\label{fig:ranges}
{\it The allowed ranges of (a, b) the spin-independent cross section 
and (c, d) the spin-dependent cross section, for (a, c) $\mu > 
0$ and (b, d) $\mu < 0$. The darker solid (black) lines show the upper 
limits on 
the cross sections obtained from $m_h$ and $b \to s \gamma$, and (where 
applicable) the lighter 
solid (red) lines show the lower limits suggested by $g_\mu - 2$ and the 
dotted (green) lines the lower limits from the relic density.
}}
\end{figure}

We see in panel (f) of Fig.~\ref{fig:decimation} that the spin-independent 
cross section for $\mu > 0$ may be rather larger for $\tan \beta = 50$ 
than for $\tan 
\beta = 10$ \cite{minmax}, as shown in panel (a). This analysis is 
extended 
in panels (a) and (c) of Fig.~\ref{fig:ranges} to all the values $8 
< \tan \beta \le 55$ (below $\tan \beta \simeq 8$ it is not possible to
satisfy both the Higgs mass and $g-2$ constraints \cite{susyg-2,ENO}, and
above
$\tan \beta \simeq 55$  we no longer find consistent CMSSM 
parameters), and we find overall that
\begin{eqnarray}
2 \times 10^{-10}~{\rm pb} \la & \sigma_{SI} & \la 6 \times 10^{-8}~{\rm
pb},
\\ 2 \times 10^{-7}~{\rm pb} \la & \sigma_{SD} & \la  10^{-5}~{\rm
pb},
\label{xsecrangeallp}
\end{eqnarray}
for $\tan \beta \le 55$ and $\mu > 0$. As we see in panels (a) and (c) 
of Fig.~\ref{fig:ranges}, for $\mu > 0$, $m_h$ provides the most 
important upper limit 
on the cross sections for $\tan \beta < 23$, and $b \to s \gamma$ for 
larger $\tan \beta$, with $g_\mu - 2$ always providing a more stringent 
lower limit than the relic-density constraint. 
The relic density constraint shown is evaluated at the endpoint of the
coannihilation region. At large $\tan \beta$, we have not considered 
moving far out into the
Higgs funnels or the focus-point regions, as their locations are very 
sensitive to input parameters and calculational details~\cite{EO}. In
the case
$\mu < 0$, there  is no lower limit on the spin-independent cross
section, for the reasons  discussed earlier. We find
\begin{eqnarray}
 & \sigma_{SI} & \la  2 \times 10^{-10}~{\rm pb}, \\
2 \times 10^{-8}~{\rm pb} \la & \sigma_{SD} & \la  2 \times 10^{-6}~{\rm
pb}
\label{xsecrangealln}
\end{eqnarray}
for $\mu < 0$ and $5 < \tan \beta \le 35$ (below $\tan \beta \simeq 5$ it
is not possible to satisfy both the Higgs mass and relic density
constraints
\cite{EGNO}, and above
$\tan \beta \simeq 35$ we no longer find 
consistent CMSSM parameters), with the upper limits being imposed 
by $m_h$ for $\tan \beta < 12$ and by
$b \to s \gamma$ for larger $\tan \beta$, as seen in panels (b) and (d) 
of Fig.~\ref{fig:ranges}. The relic-density constraint 
imposes an interesting lower limit on $\sigma_{SD}$, but not on 
$\sigma_{SI}$, as discussed above.  Again, requiring $m_h
< 117$~GeV would impose a lower limit $\sigma_{SD} \ga 3 \times
10^{-7}~{\rm pb}$, and since a 117 GeV bound would cut out the
cancellation region, we can obtain a lower bound on the spin-independent
cross section,
$\sigma_{SI} \ga 10^{-11}$ for
$\tan
\beta = 35$ and 
$\mu < 0$.

We conclude that the available experimental constraints on CMSSM model 
parameters greatly restrict the allowed ranges of elastic scattering cross 
sections for supersymmetric dark matter. Upper limits are imposed on both 
$\sigma_{SI}$ and $\sigma_{SD}$ by both the LEP Higgs constraint and $b 
\to s \gamma$. If one takes at face value the $g_\mu - 2$ constraint, in 
addition to requiring $\mu > 0$, it also imposes lower limits on both 
$\sigma_{SI}$ and $\sigma_{SD}$, providing experiments with a plausible 
sensitivity to aim for. On the other hand, if one drops the $g_\mu - 2$ 
constraint and tolerates $\mu < 0$, there is no useful lower limit on
$\sigma_{SI}$. A lower bound on $\sigma_{SD}$ is possible if one imposes
$m_h < 117$~GeV,  motivated by the LEP Higgs `hint'. The LEP constraints
are now stable, but the situation with 
$g_\mu - 2$ can be expected to clarify soon. If the apparent deviation 
from the Standard Model~\cite{g-2} is confirmed, direct searches for 
supersymmetric dark matter may have bright prospects, at least within the 
CMSSM framework studied here.

\
\vskip 0.5in
\vbox{
\noindent{ {\bf Acknowledgments} } \\
\noindent 
The work of K.A.O. was supported in part by DOE grant
DE--FG02--94ER--40823.}

\end{document}